\documentclass[pre,nofootinbib,twocolumn,showpacs,showkeys]{revtex4}
\usepackage{hyperref}
\begin{document}
\title{The Backwards Arrow of Time of the Coherently Bayesian Statistical
  Mechanic}
\author{Cosma Rohilla Shalizi}
\affiliation{Center for the Study of Complex Systems, 4485 Randall Laboratory, University of Michigan, Ann Arbor, MI 48109 USA}
\email{cshalizi@umich.edu}
\begin{abstract}
Many physicists think that the maximum entropy formalism is a straightforward
application of Bayesian statistical ideas to statistical mechanics.  Some even
say that statistical mechanics is just the general Bayesian logic of inductive
inference applied to large mechanical systems.  This approach identifies
thermodynamic entropy with the information-theoretic uncertainty of an (ideal)
observer's subjective distribution over a system's microstates.  In this brief
note, I show that this postulate, plus the standard Bayesian procedure for
updating probabilities, implies that the entropy of a classical system is {\em
  monotonically non-increasing} on the average --- the Bayesian statistical
mechanic's arrow of time points backwards.  Avoiding this unphysical conclusion
requires rejecting the ordinary equations of motion, or practicing an
incoherent form of statistical inference, or rejecting the identification of
uncertainty and thermodynamic entropy.
\end{abstract}
% 02.50.Tt = Inference methods (probability)
% 05.20.-y = Classical statistical mechanics
% 05.70.Ln = Nonequilibrium and irreversible thermodynamics
% 89.70.+c = Information theory
\maketitle

Recent years have seen renewed interest in connections between physics and
statistics \cite{stat-phys-and-stat-inf}.  Of particular interest, naturally,
has been the connection between statistical mechanics and statistical
inference.  The subjectivist approach to statistical mechanics, ably advocated
in recent times by Jaynes \cite{Jaynes-essays} and his school, holds that
probabilities represent degrees of belief; specifically, the probability of a
microstate is the degree to which an ideal observer should believe the system
is in that state, given the evidence available, and the entropy of the system
is that observer's uncertainty as to the microstate.  The theory governing the
coherent use of subjective probabilities is called Bayesian statistics
\cite{Bernardo-et-al-bayesian-theory}.  Jaynes, in particular, claimed that
statistical mechanics is just an application of the general logic of Bayesian
inference.  The validity of statistical mechanics would then be independent of
such tricky dynamical properties as ergodicity, mixing, etc., which on other
interpretations are vital.  For subjectivists, the intensive study of the
ergodic properties of mechanical systems is simply time wasted.

While controversial \cite{Sklar-chance}, the Bayesian vision of statistical
mechanics is powerful and appealing.  However, it has a flaw which has not been
pointed out before.  The second law says that the entropy of a closed system is
non-decreasing; this provides the arrow of time.  In \S \ref{sec:derivation}, I
prove that, equating thermodynamic entropy with subjective uncertainty,
ordinary Bayesian inference implies that entropy is {\em non-increasing} over
time, at least on average and sometimes strictly.  (\S \ref{sec:limit}
investigates the long-run behavior of the distribution under Bayesian updating;
it is ancillary to the main line of argument.)  This is completely unphysical,
so \S \ref{sec:avoiding} examines the proof's assumptions.  There are strong
arguments that any coherent use of subjective probabilities {\em must} employ
Bayesian updating.  In any case, replacing it by repeated applying the maximum
entropy principle {\em still} reverses the arrow of time.  This forces a choice
between inconsistent, {\em ad hoc} rules of statistical inference, or
abandoning the equation of physical entropy with uncertainty.

This rest of this introduction fixes notation, following \cite{Lasota-Mackey,%
  Cover-and-Thomas}, and makes explicit some innocent assumptions.

Start with a classical mechanical system with a phase space $\Gamma$; write $x$
for a point in this phase space.  $F$ is a distribution on $\Gamma$,
representing an (ideal) observer's uncertainty about the microscopic state $s$.
For simplicity, assume this distribution has a density $f$ (i.e., is absolutely
continuous with respect to Lebesgue measure on $\Gamma$).  Denote expectation
by angle brackets, so the mean of $M$ is $\langle M\rangle$; subscripts will
specify the distribution used in the expectation when necessary, i.e.
${\langle M \rangle}_{F}$ is the mean of $M$ under distribution $F$.  $\langle
M | N \rangle$ is the expectation of $M$ conditional on $N$.

The system's equations of motion lead to a discrete-time evolution operator $T$
on $\Gamma$, which I will assume is non-singular.  (Everything still works in
continuous time, but needs more symbols.)  By a slight abuse of notation, $T$
also denotes the induced Frobenius-Perron operator taking distributions on
$\Gamma$ into new distributions on $\Gamma$.  The specification of $T$ also
induces an evolution operator for observables, the Koopman operator $U$, such
that $U\phi(x) =\phi(T(x))$ for any sufficiently well-behaved ($L^{\infty}$)
function $\phi$.  From this definition, it can be seen that ${\langle U\phi
  \rangle}_{F} = {\langle \phi \rangle}_{TF}$.  (The difference between the
Frobenius-Perron and Koopman operators is like that between the Schr\"odinger
and Heisenberg pictures, respectively.)

There is, in addition to the microscopic degrees of freedom, a set of
macroscopic degrees of freedom, collectively $M$.  These observables depend
only on the present microscopic state, though possibly noisily, through some
observation density $p(M=m|X=x)$.

Write $F_0$ for the initial distribution, and $F_t$ for the distribution at
time $t$.  The distribution $F_0$ may be derived via a maximum-entropy
procedure, starting from an initial observation $M(0) = m_0$
\cite{Jaynes-essays}.  However, it really doesn't matter where $F_0$ comes
from, or what form it takes.  Finally, write $H[F]$ for the Shannon entropy of
the distribution $F$, i.e.,
\begin{eqnarray}
H[F] & \equiv & -\int_{\Gamma}{f(x) \log{f(x)} dx}
\end{eqnarray}
The information content of a random variable $X$ is defined to be the entropy
of its distribution function, which for convenience will also be written
$H[X]$; it should always be clear which is meant.  Conditional information
content, $H[X|Y=y]$, is the entropy of the conditional distribution.

\section{Derivation of the Backwards Arrow}
\label{sec:derivation}

So far, I have either been fixing notation, or making assumptions which are
common to all approaches to statistical mechanics, and so presumably innocuous.
I now make three explicit and substantive assumptions.
\newcounter{assumptions} \renewcommand{\theassumptions}{\Roman{assumptions}}
\begin{list}{\Roman{assumptions}}{\usecounter{assumptions}\setlength{\rightmargin}{\leftmargin}}
\item \label{assume:invert} The evolution operator $T$ is invertible.
\item \label{assume:condition} The probability distribution over microstates
  gets updated by the usual application of Bayes's rule, ${p(X=x|Y=y)} =
  {p(Y=y|X=x)p(X=x)/p(Y=y)}$.
\item \label{assume:entropy} The thermodynamic entropy at time $t$, $S_t$, is
  equal to $H[F_t]$.
\end{list}
These assumptions reverse the arrow of time, i.e., they make entropy
non-increasing.

Begin with the initial distribution over microstates, $F_0$.  After one time
step, this is transformed to a new distribution, $TF_0$.  From a Bayesian
perspective, this does not represent a change in our {\em knowledge} of the
system, merely keeping our predictions up to date.  (Rather than updating the
distribution, we could use the Koopman operator to update observables.) It is a
well-known consequence of assumption \ref{assume:invert} that $H[TF_0] =
H[F_0]$, i.e., that conservative dynamics are entropy-preserving \cite[Theorem
  9.3.1]{Lasota-Mackey}.  However, we now make a new measurement of the
macroscopic observable $M$, getting the value $m_1$.  Then, via assumption
\ref{assume:condition}, Bayes's rule gives us a new distribution:
\begin{eqnarray}
f_1(x) & = & \frac{p(m_1|x) Tf_0(x)}{\int_{\Gamma}{p(m_1|x)Tf_0(x) dx}}
\end{eqnarray}
Now, $f_1(x)$ is simply the density of $X_1$, conditional on $M_1=m_1$.  So
$H[F_1] = H[X_1|M_1=m_1]$.  An elementary inequality of information theory
\cite{Cover-and-Thomas} tells us that ``conditioning reduces entropy'';
specifically, $H[X|M] = \langle H[X|M=m] \rangle \leq H[X]$, with equality if
and only if $X$ and $M$ are statistically independent.  Using assumption
\ref{assume:entropy} to identify the thermodynamic entropy $S_t$ and the
Shannon information $H[F_t]$,
\begin{equation}
\langle S_1 \rangle = \langle H[F_1] \rangle \leq  H[F_0] = S_0
\label{eqn:entropy-decreases}
\end{equation}
Thus, unless the macroscopic observable is in fact merely noise, the entropy
{\em decreases} on average.  While there may be values of $m$ which are so
uninformative they increase an observer's uncertainty about the microscopic
state, on average every observation helps narrow that uncertainty.

A stronger result follows from the common idealization that observables are
{\em deterministic} functions of microscopic state, $M(x) = m$.  In this case,
$p(m|x)$ is either 0 or 1, depending on whether $M(x) = m_1$ or not.  Thus
$f_1(x) = Tf_0(x) 1_{M^{-1}(m_1)}(x)/TF_0(M^{-1}(m_1)) $, i.e., the truncation
of $TF_0$ to the part of $\Gamma$ compatible with the macroscopic observation.
Unless $M^{-1}(m_1)$ includes the entire support of $TF_0$, $F_1$ is a more
concentrated measure than $TF_0$ or $F_0$, and so the entropy has strictly
decreased, and not just on average.\footnote{The most important case where
  $\mathrm{supp}~TF_0 \subseteq M^{-1}(m_1)$ is when $M$ is a constant of the
  motion, e.g., total energy for a Hamiltonian system.  Entropy is then
  constant after the first measurement, even if the system begins arbitrarily
  far from equilibrium.}

Under repeated measurements, the entropy is non-increasing, either on average
or strictly, depending on whether the measurements are noisy or not.  (Entropy
is constant between observations.)  In the case of discrete-valued
deterministic measurements, if the measurement partition is ``generating''
\cite{Badii-Politi,beim-Graben-Atmanspacher-classical-complementarity}, then
the volume of $\Gamma$ compatible with a sequence of measurements shrinks
towards zero, and so the uncertainty, as measured by the Shannon information,
tends to $-\infty$.  This is not necessarily the case if the measurement
partition is not generating.

Note that I required nothing of the dynamics other than assumption
\ref{assume:invert}, invertibility.  In particular, I did not need chaos,
ergodicity, mixing, etc., either at the microscopic or macroscopic level.
Thermodynamic equilibrium or its absence is also irrelevant.

\subsection{Long-Run Behavior of $H[F_t]$}
\label{sec:limit}

Describing the long-run behavior of $H[F_t]$ requires explicit use of
measure-theoretic probability \cite{Pollard-users-guide}, and what's called
``Doob's martingale''.  As a measurable space, $\Gamma$ comes with a
$\sigma$-algebra of measurable sets $\mathcal{G}$.  Let $G$ be any set in
$\mathcal{G}$.  Then $1_{G}(X)$ is a random variable, indicating whether or not
$X \in G$, and ${\langle 1_{G} \rangle}_{F_0} = F(G)$, the probability of the
set $G$ under distribution $F_0$.  Let $\mathcal{M}_t = \sigma(M_1, \ldots
M_t)$, the smallest $\sigma$-algebra with respect to which all the observables
up to $M_t$ are measurable, and examine ${\langle
  1_{G}|\mathcal{M}_t\rangle}_{F_0}$, the conditional expectation of the
indicator variable for $G$.  Clearly, ${\langle
  1_{G}|\mathcal{M}_t\rangle}_{F_0} = {\langle 1_{G} \rangle}_{F_t} =
F_t(G)$. $F_t(G)$ is a martingale and converges almost surely and in mean
square to a random variable $F_{\infty}(G)$, which is the conditional
expectation of $1_G$ with respect to $\mathcal{M}_{\infty}$, the smallest
$\sigma$-algebra containing all the $\mathcal{M}_{t}$ \cite[\S
  6.6]{Pollard-users-guide}.  Thus, the conditional measures $F_t$ converge
weakly on a limit $F_{\infty}$ \cite[\S 7.1]{Pollard-users-guide}.  Thus, the
entropy of $H[F_t]$ also converges on a limiting value.  If
$\mathcal{M}_{\infty} = \mathcal{G}$, as in the generating partition case,
then, for any set $G$, $F_{\infty}(G) = 0$ or $=1$, and $H[F_{\infty}] =
-\infty$.  If $\mathcal{M}_{\infty} \subset \mathcal{G}$, then the conditional
distributions converge weakly on a distribution with a finite
entropy.\footnote{By Eq. \ref{eqn:entropy-decreases}, the sequence $H[F_t]$
  forms a supermartingale with respect to the filtration induced by the
  macro-variables $M_t$, but the conditions needed to directly apply martingale
  convergence theorems, such as $\langle |H[F_t]|\rangle < \infty$, do not
  necessarily hold.}

A somewhat more refined result is possible if we assume that the asymptotic
equipartition property of information theory holds (i.e., that the
Shannon-Macmillan-Breiman theorem applies).  This leads to estimates of the
asymptotic growth rates for likelihoods, and so for posterior probabilities in
Bayes's rule.

Suppose that the following limits exist for every $x, y \in \Gamma$:
\begin{eqnarray}
\nonumber
\lefteqn{h(x)} & &\\
& \equiv & \lim{\frac{1}{n}H[M_n|M_{n-1}, M_{n-2} \ldots M_{1}, X_0=x]}\\
& = & -\lim{\frac{1}{n}\int{d^{n}m~ p(m_1^n|x)\log{p(m_{1}^{n}|x)}}}\\
\nonumber
\lefteqn{d(x,y)} & & \\
 & \equiv & \lim{\frac{1}{n}\int{d^{n}m~ p(m_1^n|y)\log{\frac{p(m_1^n|x)}{p(m_1^n|y)}}}}
\end{eqnarray}
where $p(m_1^n|x)$ abbreviates $p(M_1 = m_1, M_2 = m_2, \ldots M_n = m_n|X_0 =
x)$.  The quantity $h(x)$ is the macroscopic entropy rate at $x$ (not to be
confused with the {\em microscopic} rate of entropy production
\cite{Ruelle-smooth-dynamics}).  $d(x,y)$ is the macroscopic relative entropy
rate, or Kullback-Leibler divergence rate, between $x$ and $y$.  Note that
$d(x,y) \geq 0$, and that $d(x,y)=0$ if and only if
${p(M_n=m_n|M_1^{n-1}=m_1^{n-1},X_0=y)}$ and ${p(M_n=m_n|M_1^{n-1},X_0 =x)}$
converge for almost all $m_1^n$ \cite{Gray-entropy}.  Further assume that the
asymptotic equipartition property \cite{Gray-entropy} holds, so that, if $X_0 =
y$, then for $F_0$-almost-all $x$
\begin{eqnarray}
\lim{-\frac{1}{n}\log{p(m_1^n|x)}} & = & h(y) + d(x,y)
\label{eqn:aep}
\end{eqnarray}
almost surely.  An immediate corollary of Eq.\ \ref{eqn:aep} is that
\begin{eqnarray}
\log{p(m_1^n|x)} & = & -nh(y) - nd(x,y) + g(x,y,m_1^n)
\label{eqn:extended-aep}
\end{eqnarray}
where $g$ is a random quantity which is $o(n)$ almost surely.

Write Bayes's rule with $n$ observations in logarithmic form, and substitute
in Eq.\ \ref{eqn:extended-aep} (assuming $f_0(x) > 0$):
\begin{eqnarray}
\lefteqn{\log{\frac{f_n(x)}{f_0(x)}} = \log{p(m_1^n|x)} - \log{{\left\langle p(m_1^n|x)\right\rangle}_{F_0}}} & & \\
\nonumber
 & = & -nh(y) - nd(x,y) + g(x,y,m_1^n) \\
& & - \log{{\left\langle e^{-nh(y) - nd(x,y) + g(x,y,m_1^n)}\right\rangle}_{F_0}}\\
\nonumber
& = & -nh(y) - nd(x,y) + g(x,y,m_1^n)\\
& & -\log{e^{-nh(y)}{\left\langle e^{-nd(x,y)} e^{g(x,y,m_1^n)}\right\rangle}_{F_0}}\\
\nonumber
& = & -nd(x,y) + g(x,y,m_1^n) - \gamma(y,m_1^n)\\
& & - \log{{\left\langle e^{-nd(x,y)}\right\rangle}_{F_0}}
\label{eqn:break-in-log-bayes}
\end{eqnarray}
$\gamma$ is another $o(n)$ random quantity; for later use, set $\eta(x,y,m_1^n)
= g(x,y,m_1^n) - \gamma(y,m_1^n)$.
\begin{eqnarray}
\nonumber
\lefteqn{\log{{\left\langle e^{-nd(x,y)}\right\rangle}_{F_0}}} & &\\
& = & \log{\int{dx f_0(x) {\left(e^{-d(x,y)}\right)}^n}} \\
& = & \log{{\left({\left\|e^{-d(x,y)}\right\|}^{n}_{F_0}\right)}^n}\\
& = & n\log{{\left\|e^{-d(x,y)}\right\|}^{n}_{F_0}}\\
& = & -n \overline{d_n(y)}
\end{eqnarray}
The last line defines $\overline{d_n(y)}$, and
${\left\|\phi(x)\right\|}^{n}_{F} \equiv {\left({\left\langle {\left|\phi(x)
      \right|}^{n} \right\rangle}_{F}\right)}^{1/n}$ is the $L^n$ norm of the
function $\phi$ with respect to the measure $F$ \cite{Pollard-users-guide}.
The latter is non-decreasing in $n$, and its limit is the essential supremum of
the function.  That is, ${\left\|e^{-d(x,y)}\right\|}^{\infty}_{F_0}$ is the
smallest $u$ such that $u \geq e^{-d(x,y)}$ for all $x$, except on a set of
$F_0$-probability zero.  It follows that $d_{\infty}(y)$ is the essential
infimum of $d(x,y)$, and so $d_{\infty}(y) \leq d(x,y)$ everywhere except on a
set of $F_0$-probability zero.  Since $d(x,y) \geq 0$, we can be sure that
$\overline{d_{\infty}(y)}$ is at least zero.  Since $d(y,y) = 0$, if $f_0$ is
positive in every sufficiently small neighborhood of $y$, we will have
$\overline{d_{\infty}(y)} = 0$.  Since the procedures used to construct prior
distributions for statistical mechanics generally give non-vanishing weight to
all physically accessible regions of the phase space, this last assumption is
reasonable, and so set $\overline{d_{\infty}(y)} = 0$.

Substituting back in to Eq.\ \ref{eqn:break-in-log-bayes},
\begin{eqnarray}
\log{\frac{f_n(x)}{f_0(x)}} & = & -nd(x,y) + \eta(x,y,m_1^n) + n\overline{d_n(y)} \\
& = & n(\overline{d_n(y)} - d(x,y)) + \eta(x,y,m_1^n)
\end{eqnarray}
Taking the limit as $n\rightarrow\infty$,
\begin{eqnarray}
\lim{\frac{1}{n}\log{\frac{f_n(x)}{f_0(x)}}} & = & \overline{d_{\infty}(y)} - d(x,y)\\
& = & -d(x,y)
\end{eqnarray}
Asymptotically, therefore, $f_n(x)$ shrinks exponentially fast towards zero,
unless $d(x,y) = 0$.  Setting $D(y) = \left\{x\left| d(x,y) =
0\right.\right\}$, we see that $F_{\infty}(D(y)) = 1$, and so $H[F_{\infty}]$
is at most the logarithm of the volume of $D(y)$.

\section{Ways to Avoid This Result}
\label{sec:avoiding}

Since, in reality, thermodynamic entropy is monotonically non-decreasing, at
least one of the assumptions leading to Eq.\ \ref{eqn:entropy-decreases} must
be wrong.

\subsection{Assumption \ref{assume:invert}: Invertible Dynamics}

Denying assumption \ref{assume:invert} ``has all the advantages of theft over
honest toil'' \cite{Russell-intro-to-math-phil}.  The problem of the
foundations of statistical mechanics is precisely that of deriving macroscopic
irreversibility from microscopically-reversible dynamics, and the point of the
Bayesian approach was to do so with making detailed assumptions about those
dynamics.  That said, crime may not pay; the conditions needed to get $H[TF]
\geq H[F]$ impose highly non-trivial restrictions on the dynamics
\cite{Mackey-times-arrow}.  Not only are conservative Hamiltonian dynamics
ruled out, but so is any system of ordinary differential equations.  What is
really needed is that $H[TF|M] \geq H[F]$, and it is hard to see why the
microscopic dynamics should {\em always} produce more than enough entropy to
off-set the information provided by whichever macroscopic observables we happen
to choose.  But without such cancellation, watching a pot {\em closely enough}
will keep it from boiling.

\subsection{Assumption \ref{assume:condition}: Bayesian Updating}

Explicitly or not, most advocates of Bayesian statistical mechanics reject
assumption \ref{assume:condition} \cite{Sklar-chance,Uffink-constraint-rule}.
For instance, Jaynes put forward the following derivation of the second law
\cite{Jaynes-essays}: Start with an initial observation of an observable, $M_0
= m_0$.  Confine ourselves to distributions $\rho$ which have ${\langle M
  \rangle}_{\rho} = m_0$; call the set of such distributions $C_0$.  Let the
member of $C_0$ with the highest entropy be $J_0$; we select this as our
initial distribution.  Now let it evolve forward in time, giving $TJ_0$; by
assumption \ref{assume:invert}, $H[J_0] = H[TJ_0]$.  The time evolution leads
to a certain value for the observable, $m_1 = {\langle M \rangle}_{T J_0}$.
Now consider the class of distributions $C_1$ with ${\langle M \rangle}_{\rho}
= m_1$; let the maximum entropy member of this class be $J_1$.  Since $TJ_{0}
\in C_1$, it follows that $H[J_1] \geq H[T J_0] = H[J_0]$.  Jaynes then
identifies $S_1$ with $H[J_1]$, i.e., he updates the distribution by
re-applying the maximum entropy principle, using only the {\em current}
observation, rather than by applying Bayes's rule.

There are good reasons to doubt the wisdom of using probability as a measure of
degree of belief \cite{Mayo-error}.  But if you {\em are} going to do that,
then the Bayesian way is the right way to do so, and you need to use
conditioning.  Failure to do so is incoherent, as the well-known ``Dutch Book''
arguments show.  (\cite{Hacking-intro-to-prob} gives a clear introduction; see
\cite{Bernardo-et-al-bayesian-theory} for details.)  In particular, in the
formally very similar problem of nonlinear filtering \cite{Ahmed-filtering},
application of Bayes's rule is demonstrably optimal, and forgetting all earlier
observations is {\em not}, regardless of whether one interprets probability
subjectively.\footnote{The doubts raised by
  \cite{Bacchus-et-al-against-conditionalization} about inter-temporal updating
  are not relevant.  There's no time lapse between $TF_0$ and $F_1$, just the
  addition of the information that $M_1=m_1$.}

Even if Bayesian statistical mechanics are free to not use conditioning, they
still get a backwards arrows of time.  A consistent use of the principle of
maximum entropy, given the two observations $M_0 = m_0$ and $M_1 = m_1$, would
go as follows.  First, restrict ourselves to distributions $\rho$ which satisfy
both the constraints ${\langle M \rangle}_{\rho} = m_0$ and ${\langle M \rangle
}_{T \rho} = m_1$.  It is awkward to have one constraint on $\rho$ and another
on $T\rho$; using the Koopman operator, we can turn the latter into a
constraint on $\rho$ as well, ${\langle UM \rangle}_{\rho} = m_1$.  Let us
write $C_{01}$ for the class of distributions satisfying these two constraints.
Those satisfying the first constraint are the class we called $C_0$ above, and
those satisfying the second constraint form a subclass of the set we called
$C_1$ above.  Hence $C_{01} \subseteq C_0 \cap C_1$.  Then the maximum entropy
principle tells us to pick the distribution $J_{01}$ given by
\begin{eqnarray}
J_{01} & = & \mathrm{arg}\max_{\rho \in C_{01}}{H[\rho]}
\end{eqnarray}
Since $C_{01} \subseteq C_0$ and $C_{01} \subseteq C_1$, it is clear that
$H[J_{01}] \leq \min{H[J_0], H[J_1]}$.  Thus, updating our distribution by
maximizing entropy, rather than conditioning, {\em still} reverses the arrow of
time: by assumption \ref{assume:entropy}, $S_1 = H[T J_{01}]$, and by
assumption \ref{assume:invert}, $H[T J_{01}] = H[J_{01}] \leq H[J_0] = S_0$.

To avoid getting the direction of the arrow of time backwards, the Bayesian or
Jaynesian statistical mechanic must ignore the {\em known} prior history, a
procedure quite without statistical justification.  By use of the operator $U$,
constraints on a single observable over multiple times can be converted into
constraints on multiple observables at a single time, which we are normally
told must {\em all} be incorporated into the distribution $F$.  There does not
seem to be any reason why it should be legitimate to take $M$ as a constraint
in the maximum entropy procedure, but not $UM$.  Worse yet, under some
circumstances subjectivists (e.g., Jaynes, in his discussion of spin-echo
experiments \cite{Jaynes-essays,Sklar-chance}) have been explicit about needing
to incorporate historical information in order to avoid unphysical predictions.

Whether we update via conditioning, or by applying the maximum entropy
principle, we get unphysical results for the entropy, and can avoid them only
by inconsistency about whether historical data counts, or, equivalently,
whether all observables must inform the postulated distribution.  One might
argue that, for most systems of interest, the distribution obtained from
applying the maximum entropy principle only to ordinary observables at the
current time leads to nearly the same {\em predictions} as the coherent
procedures, but the former is much easier to calculate than the latter,
particularly if the dynamics are very irregular.  In such a case, the
complexity of computing the $n^{\mathrm th}$ iterate of the evolution operator,
$T^n$, may grow rapidly with $n$, so that a computationally-limited agent,
acting under time pressure, might prefer an approximation which neglects
historical data to an exact but intractable update.  The validity of such an
approximation would depend on the ergodic properties of the dynamics (e.g.,
mixing), and it is precisely to avoid such dependence that the Bayesian
approach was introduced.  Worse, it would lead to a novel kind of Maxwell's
demon, a purely passive observer who can make the entropy decrease during a
time interval which depends on the observer's processing speed and the
time-complexity of computing $T^n$.  In any event, none of this would explain
why the thermodynamic entropy should match the entropy of {\em this}
approximate distribution.

\subsection{Assumption \ref{assume:entropy}: Thermodynamic Entropy Is Subjective Uncertainty}

Assumption \ref{assume:entropy} is that thermodynamic entropy $S$ is the
information-theoretic uncertainty $H[F]$.  Denying this seems to me a
completely satisfactory option.  Macroscopically, entropy is defined by its
relations to the observables of heat and temperature.  Microscopically,
assuming the usual representation of phase space, entropy is the logarithm of
the volume in phase space compatible with the current macroscopic state
\cite{Lebowitz-central-issues,Sewell-emergent-macrophysics,%
  Gross-topology-of-microcannonical}; more generally, it is the logarithm of
the measure of that region \cite{Albert-time-and-chance,What-is-a-macrostate}.

Rejection of assumption \ref{assume:entropy} is perfectly compatible with
accepting a Bayesian, subjectivist interpretation of probability.

\section{Conclusion}

A backwards arrow of time follows directly from the combination of assumptions
I, II and III, at least one of which must therefore be rejected.  Rejecting
assumption \ref{assume:invert}, invertible microphysical dynamics, entails
considerable modification of basic physics.  Rejecting assumption
\ref{assume:condition}, updating subjective probabilities via Bayes's rule, is
actually insufficient; one must also reject the principle of maximum entropy,
or at the very least apply it in an incoherent way, sometimes taking into
account all observational constraints, sometimes not.  Rejecting assumption
\ref{assume:entropy}, the identification of thermodynamic entropy with the
Shannon information $H[F]$, seems to lead to the least trouble.  I do not
pretend that only one choice among these alternatives is defensible, but some
choice is necessary.

{\em Acknowledgments.}  This work was supported by a grant from the James
S. McDonnell Foundation.  Thanks to Wolfgang Beirl, Fred Boness, Milan
Cirkovic, D. H. E. Gross, Aleks Jakulin, Cris Moore and Eric Smith for
comments, and to Eric for the remark which led me to write this.

\bibliography{locusts}
\end{document}